\documentstyle[fleqn,twoside]{article}

\def\be{\begin{equation}}
\def\ee{\end{equation}}
\def\ai{\'{\i}}

\def\tp1{\tilde p_1}
\def\tp2{\tilde p_2}
\def\tq1{\tilde q^1}
\def\tq2{\tilde q^2}

\newcommand{\Title}[1]{\noindent {\uppercase{\Large #1}} \\}
\newcommand{\Authors}[4]{\noindent
  {\large\bf #1\dag\ #2\ddag}\medskip\begin{description}
  \item[\dag]{\it #3} \item[\ddag]{\it #4}\end{description}}
\newcommand{\Abstract}[1]{\vskip 2mm \begin{center}
  \parbox{16.4cm}{\small\noindent #1} \end{center}\medskip}
\newcommand{\foom}[1]{\protect\footnotemark[#1]}
\newcommand{\email}[2]{\footnotetext[#1]{e-mail: #2}
  \addtocounter{footnote}{1}}

\makeatletter
\renewcommand{\@oddhead}{\raisebox{0pt}[\headheight][0pt]{%
   \vbox{\hbox to\textwidth{{\protect\small\it %
   The Problem of Time and Gauge Invariance -- II}%
   \hfil \rm \thepage \strut}\hrule}}}
\renewcommand{\@evenhead}{\raisebox{0pt}[\headheight][0pt]{%
   \vbox{\hbox to\textwidth{\thepage \hfil {\protect\small\it %
   T. P. Shestakova and C. Simeone} \strut}\hrule}}}
\renewcommand{\section}{\@startsection{section}{1}{0pt}%
  {-3.5ex plus -1ex minus -.2ex}{2.3ex plus .2ex}%
  {\large\bf\protect\raggedright}}

\renewcommand{\subsection}{\@startsection{subsection}{2}{0pt}%
  {-3ex plus -1ex minus -.2ex}{1.4ex plus .2ex}%
  {\normalsize\bf\protect\raggedright}}

\makeatother

\begin{document}
\twocolumn[
\thispagestyle{empty}
\bigskip

\Title{THE PROBLEM OF TIME AND GAUGE INVARIANCE \\[5pt]
       IN THE QUANTIZATION OF COSMOLOGICAL MODELS. \\[5pt]
       II. RECENT DEVELOPMENTS IN THE PATH INTEGRAL APPROACH}

\Authors{T. P. Shestakova\foom 1}
        {and C. Simeone\foom 2}
        {Department of Theoretical and Computational Physics,\\
         Rostov State University, Sorge Str. 5, Rostov-on-Don, 344090, Russia}
        {Departamento de F\'{\i }sica,
         Facultad de Ciencias Exactas y Naturales Universidad de Buenos Aires,\\
         Ciudad Universitaria Pabell\'{o}n I - 1428, Buenos Aires, Argentina\\
         and Instituto de Astronom\ai a y F\ai sica del Espacio
         C.C. 67, Sucursal 28 - 1428 Buenos Aires, Argentina}

\Abstract
{The paper is the second part of the work devoted to the problem of
time in quantum cosmology. Here we consider in detail two
approaches within the scope of Feynman path integration scheme: The
first, by Simeone and collaborators, is gauge-invariant and lies
within the unitary approach to a consistent quantization of
gravity. It is essentially based on the idea of deparametrization
(reduction to physical degrees of freedom) as a first step before
quantization. The other approach by Savchenko, Shestakova and
Vereshkov is rather radical. It is an attempt to take into account
peculiarities of the Universe as a system without asymptotic states
that leads to the conclusion that quantum geometrodynamics
constructed for such a system is, in general, a gauge-noninvariant
theory. However, this theory is shown to be mathematically
consistent and the problem of time is solved in this theory in a
natural way.}]

\email 1 {shestakova@phys.rsu.ru}
\email 2 {csimeone@df.uba.ar}
\date{}

\section{Introduction}

In Part I of our work \cite{shsi} we have considered most
representative approaches to the well-known problem of time in
quantum cosmology which lie in the framework of canonical
quantization. Unfortunately, most of these proposals can be applied
only to restricted classes of models. The most interesting and
promising approaches which go beyond the minisuperspace
approximation are the proposals by Brown and Kucha\v r \cite{ku95}
and by Barvinsky and collaborators \cite{ba86a,ba86b,ba87,ba93},
the later having been formulated mainly making use the Feynman path
integral formalism.

The main object of the path integral approach \cite{di33,fey48,fey65} is
a transition amplitude between two states which is obtained as the
sum over all histories of the exponential of the action. For a
constrained system, divergences yielding from the overcounting of
paths in phase space which are physically equivalent should be
avoided by imposing gauge conditions that select one path from each
class of equivalence \cite{fapo67,fasl80}. In its phase space form
the propagator then reads
\begin{eqnarray}
\langle q^i_2|q^i_1\rangle & = & \nonumber\\
& &\hspace{-1.6cm}\int Dq^i Dp_i DN \delta(\chi) |[\chi,{\cal H}]|
\exp
\left(i S[q^i,p_i,N]\right).
\end{eqnarray}
Here $\chi=0$ is a gauge fixing function and
$|[\chi,{\cal H}]|$ is the Faddeev -- Popov determinant, which
makes the result independent of the gauge choice. Admissible gauge
conditions are those which can be reached from any path by
performing a gauge transformation which is compatible with the
symmetries of the action.

Because in gravitational dynamics the Hamiltonian generates the
evolution and also acts as a generator of gauge transformations, it
is natural to think that a time could be defined by means of gauge
fixation, so that the resulting non divergent amplitudes would
include a clear notion of evolution. But the problem arises that
gauges defining a time in terms of the canonical variables are the
so-called canonical gauges, which can be imposed only if the
constraints are linear in the canonical momenta, and the
Hamiltonian constraint in the gravitational action is quadratic in
the momenta \cite{te82}. This seemed to appear as an obstacle for a
program of deparametrization based on this idea; however, we shall
show in Section 2 that for a class of cosmological models this can
be solved by associating to them an ordinary gauge system (that is
a system with constraints linear in the momenta), so that fixing
the gauge in the gauge system defines a time for the corresponding
minisuperspace \cite{si99}.

In Section 2 we shall follow Simeone and collaborators
\cite{fesi97,si98,si99,si00,desi99b,gisi01a,gisi01b,gisi02,si02,si02b}.
The connection between fixing an admissible gauge condition and a
time definition will be considered in detail in Section 2.1. In
Section 2.2 we shall describe a special canonical transformation
that gives rise to an action for a system with a zero Hamiltonian
and a constraint which is linear in the momenta. On this way we
face the problem of observations mentioned in Part I of our work:
new canonical variables appear to be conserved quantities since
they commute with a new Hamiltonian. So we need another canonical
transformation which leads to a time-dependent Hamiltonian. This
will be discussed in Section 2.3. We shall come to a formulation in
terms of true degrees of freedom in what we call the reduced phase
space. It allows us to define a transition amplitude through a path
integral by the usual Faddeev -- Popov procedure in Section 2.4.
Examples will be given in Section 2.5.

In the approaches by Barvinsky and by Simeone and collaborators
time is introduced into the theory by means of a time-dependent
gauge condition. In Section 3 it will be shown that time may appear
as a consequence of breaking down gauge invariance of the theory,
even if the gauge condition does not depend on time. In the
approach presented in papers by Savchenko, Shestakova and Vereshkov
\cite{ssv99, ssv00, shest99, ssv01a, ssv01b} the authors argued
that the breakdown of gauge invariance is inevitable since the
Universe as a physical system does not possess asymptotic states.
It prevents from imposing asymptotic boundary conditions which
eventually ensure gauge invariance. This will be discussed in
Section 3.1. In Section 3.2 dynamics of a simple minisuperspace
model in extended phase space will be constructed and its quantum
version will be explored in Section 3.3. At last, in Section 3.4 we
shall touch upon an intriguing question if irreversibility of time
could be related to nontrivial topology of the Universe.

\section{Path integral quantization of minisuperspaces as ordinary gauge
systems}

In this section we shall review our procedure for associating an
ordinary gauge system to a minisuperspace model, which allows to
effectively deparametrize the minisuperspace and to obtain a
consistent path integral quantization. The analogy between gauge
transformations and dynamical evolution reflected in equations
\begin{equation}
\label{evol}
{\frac{dq^{i}}{d\tau }}=N_\mu[q^{i},{\cal H^\mu}],\ \ \ \ \ \ \ \ \
{\frac{dp_{i}}{
d\tau }}=N_\mu[p_{i},{\cal H^\mu}]
\end{equation}
and
\begin{eqnarray}
\label{gauge}
\delta _{\epsilon }q^{i} & = &\epsilon_\mu (\tau )[q^{i},{\cal H^\mu}],\nonumber\\
\delta
_{\epsilon }p_{i}& = &\epsilon_\mu (\tau )[p_{i},{\cal H^\mu}],\nonumber\\ \delta
_{\epsilon }N_\mu & = &{\frac{\partial \epsilon_\mu (\tau )}{\partial \tau }}-
u_\mu^{\nu\rho}\epsilon_\rho N_\nu
\end{eqnarray}
is the basic idea leading to the reduction procedure identifying
physical degrees of freedom and time. However, because of the lack
of gauge invariance at the end points in the action of gravitation
resulting from the quadratic form of the Hamiltonian constraint,
admissible gauges would not be of the canonical form $\chi
(q^i,p_i,\tau)=0$; hence in order to perform the deparametrization,
we shall introduce a reformulation of the theory leading to a
globally gauge invariant action \cite{si99}.

\subsection{Gauge fixation and deparametrization}

Admissible gauge conditions are those which can be reached from any
path by means of gauge transformations leaving the action
unchanged, and such that only one point of each orbit is on the
manifold defined by $\chi=0$. This requires to analyse the
possibility of the {\it Gribov problem} \cite{gribov,hete92}, that
is, that depending of the form of the orbits and on the topology of
the constraint surface, it may be difficult to intersect it with a
gauge condition which is crossed by each orbit only once.

If it is possible to perform a canonical transformation $(q^i,p_i)$
$\to$ $(Q^i,P_i)$ such that the Hamiltonian ${\cal H}$ is matched
to one of the new momenta, in terms of the new variables the action
functional would include a constraint which is linear and
homogeneous in the momenta. This is equivalent to say that the
canonical variables $(Q^i,P_i)$ describe an ordinary gauge system,
so that canonical gauges $\chi(Q^i,P_i,\tau)=0$
would  be admissible.

The condition that a gauge transformation moves a point of an orbit
off the surface $\chi=0$ is fulfilled if
\begin{equation}
 [\chi,{\cal H}]\neq 0.\label{31}
\end{equation}
Now, as $Q^0$ and $P_0$ are conjugated variables,
\begin{equation}
[Q^0,P_0]=1\label{33}
\end{equation}
and if we  identify ${\cal H}\equiv P_0$, then a gauge condition of the form
\begin{equation}
\chi\equiv Q^0-T(\tau)=0\label{34}
\end{equation}
with $T$ a monotonous function is a good choice. Equation
(\ref{31}) only ensures that the orbits are not tangent to the
surface $\chi=0$; however, as (\ref{34}) defines a plane
$Q^0=constant$ for each $\tau$, if at any $\tau$ any orbit was
intersected more than once (then yielding Gribov copies) at another
$\tau$ it should be $[\chi, P_0]=0$. Therefore this gauge fixation
procedure avoids the Gribov problem \cite{si98}.

The connection with the identification of time is the following: as
we have already seen, for a parametrized system whose canonical
variables are $(q^i,p_i)$, a global phase time $t(q^i,p_i)$ is a
function fulfilling \cite{ha86}
\begin{equation}
[t,{\cal H}]>  0.\label{35}
\end{equation}
As the Poisson bracket is invariant under a canonical
transformation, from (\ref{33}) and (\ref{35}) it follows that a
global phase time can be defined for a minisuperspace by imposing
on its associated gauge system a gauge condition in terms of the
coordinate $Q^0$. In other words, {\it a gauge choice for the gauge
system defines a particular foliation of spacetime for the
corresponding cosmological model} \cite{si99}. If a gauge choice
avoiding the Gribov ambiguity can be found, then a definition of
time which is good everywhere is obtained. A transformation such
that ${\cal H}=P_0$ can always be found locally; in the next
paragraphs we shall show how a canonical transformation which works
in the whole phase space can be found.

\subsection{Gauge-invariant action for a minisuperspace}

Here we shall review our procedure to obtain a gauge-invariant
action for cosmological models whose Hamiltonian constraint is
such that a solution for its associated  $\tau-$independent Hamilton--Jacobi
equation can be found. Consider a complete solution \cite{dau60}
$W(q^i,\alpha_\mu,E)$ of the Hamilton--Jacobi equation
\begin{equation}
H\left(q^i,{\partial W\over\partial q^i}\right)=E
\end{equation}
where $H$ is not necessarily the original Hamiltonian constraint
but it can be a scaled Hamiltonian, that is $H={\cal F}^{-1}{\cal
H}$ with ${\cal F}$ a positive definite function of $q^i$. If $E$
and the integration constants $\alpha_\mu$ are matched to the new
momenta $\overline P_0$ and $\overline P_\mu$ respectively, then
$W(q^i,\overline P_i)$ turns to be the generator function of a
canonical transformation $(q^i,p_i) \to (\overline Q^i,\overline
P_i)$ defined by the equations
\begin{equation}
p_i={\partial W\over\partial q^i},\ \ \ \ \overline Q^i={\partial W\over\partial
\overline P_i},\ \ \ \  \overline K=N\overline P_0=NH \label{38}
\end{equation}
where $\overline K$ is a new Hamiltonian. The new coordinates and
momenta verify $$[\overline Q^{\mu},\overline P_0]=[\overline
Q^{\mu},H]=0$$
$$[\overline P_{\mu},\overline P_0]=[\overline P_{\mu},H]=0$$
$$[\overline Q^{0},\overline P_0]=[\overline Q^{0},H]=1.$$
The variables $(\overline Q^\mu,\overline P_\mu)$ are then {\it
observables}: they commute with the constraint, so that they are
gauge-invariant. The resulting action
\begin{equation}
\overline S[\overline Q^i,\overline P_i,N]=\int_{\tau_1}^{\tau_2}\left(\overline
P_i{d\overline Q^i\over d\tau}-N\overline P_0\right) d\tau
\end{equation}
describes a system with a zero true Hamiltonian and a constraint
which is linear and homogeneous in the momenta (hence canonical
gauges would be admissible in a path integral with this action).
The action $\overline S$ is related with $S$ by
\begin{eqnarray}
\overline S[q^i,p_i,N] &  = &  \int_{\tau_1}^{\tau_2}\left(p_i{dq^i\over d\tau}-
NH\right) d\tau\nonumber\\
& & \hspace{-1.2cm}\mbox{} +\left[\overline Q^i(q^i,p_i)\overline P_i(q^i,p_i)-W(q^i,\overline
P_i)\right]_{\tau_1}^{\tau_2},
\end{eqnarray}
so that the gauge-invariant action $\overline S$ differs from the
original action $S$ in end point terms \cite{htv92}. These terms do
not modify the dynamics, as they can be included in the action
integral as a total derivative with respect to the parameter
$\tau$.

\subsection{Time and true degrees of freedom}

The observables $\overline Q^\mu$ and $\overline P_\mu$ are
conserved quantities, because they commute with $\overline
K=N\overline P_0$. This makes impossible the characterization of
the dynamical trajectories of the system by an arbitrary choice of
$\overline Q^\mu$ at the end points $\tau_1$ and $\tau_2$. To
obtain a set of observables such that the choice of the new
coordinates is enough to characterize the dynamical evolution, non
conserved variables must be defined, and a new $\tau-$dependent
transformation leading to a non null Hamiltonian must be
introduced.

Let us consider the canonical transformation generated by
\begin{equation}
F(\overline Q^i,P_i,\tau)=P_0\overline Q^0+f(\overline Q^\mu ,P_\mu ,\tau),
\end{equation}
which leads to
\begin{eqnarray}\overline P_0 & = &{\partial F\over\partial\overline
Q^0}=P_0=H\nonumber\\  \overline P_\mu & = & {\partial F\over \partial\overline
Q^\mu}={\partial f\over \partial\overline Q^\mu},\nonumber\\
 Q^0 & = & {\partial F\over \partial P_0}=\overline Q^0,\nonumber\\  Q^\mu & = &
{\partial F\over\partial P_\mu}={\partial f\over\partial P_\mu}.\end{eqnarray}
The generator $f$ defines a canonical transformation in what we
call the reduced phase space, which corresponds to the true degrees
of freedom of the theory. The coordinates and momenta
$(Q^\mu,P_\mu)$ are observables because
$$[Q^\mu,P_0]=[P_\mu,P_0]=0,$$
but they are not conserved quantities, because their evolution is
determined by the non zero Hamiltonian
\begin{equation}
K=NP_0+{\partial f\over\partial\tau}=NH+{\partial f\over\partial\tau}.
\end{equation}
Indeed,
\begin{eqnarray}
{dQ^\mu\over d\tau} & = & {\partial K\over\partial P_\mu}  =
{\partial^2\over\partial\tau\partial P_\mu}f(\overline Q^\mu(Q^\mu,P_\mu),P_\mu,
\tau)\nonumber\\
{dP_\mu\over d\tau} & = &  -{\partial K\over\partial Q^\mu}  =  -
{\partial^2\over\partial\tau\partial Q^\mu}f (\overline
Q^\mu(Q^\mu,P_\mu),P_\mu, \tau),\nonumber\\
& &
\end{eqnarray}
so that
\begin{equation}
h(Q^\mu,P_\mu,\tau)\equiv{\partial\over\partial\tau} f(\overline
Q^\mu(Q^\mu,P_\mu),P_\mu, \tau)
\end{equation}
is a true Hamiltonian for the reduced system (below we shall give a
prescription to choose $f$). For the coordinate conjugated to $P_0$
we have
\begin{equation}
{dQ^0\over d\tau}=[Q^0,K]=N[Q^0,P_0]=N.
\end{equation}

The transformation $(\overline Q^i,\overline P_i)\to (Q^i,P_i)$
yields additional end point terms of the form
$$\left[Q^\mu P_\mu-f({\overline Q}^\mu(Q^\mu,P_\mu),P_\mu,
\tau)\right]_{\tau_1}^{\tau_2}.$$
The gauge-invariant action resulting from the two successive
canonical transformations $(q^i,p_i) \to (\overline Q^i,\overline
P_i) \to (Q^i,P_i)$ is
\begin{equation}
{\cal S}[Q^i,P_i,N]=\int_{\tau_1}^{\tau_2} \left( P_i{dQ^i\over d\tau}-NP_0-
{\partial f\over\partial\tau}\right) d\tau
\label{cals}\end{equation}
and in terms of the original  variables it includes end point terms,
\begin{eqnarray}
{\cal S}[q^i,p_i,N] &  = & \int_{\tau_1}^{\tau_2}\left( p_i{dq^i\over d\tau }-
NH\right) d\tau\nonumber\\
 & & \hspace{-2.7cm} +
\left[ \overline Q^i\overline P_i -W(q^i,\overline P_i)+Q^\mu P_\mu-f(\overline
Q^\mu ,P_\mu ,\tau)\right]_{\tau_1}^{\tau_2},
\end{eqnarray}
where $\overline Q^i,\overline P_i, Q^\mu$ and $ P_\mu$ must be
written in terms of $q^i$ and $p_i$. The action ${\cal
S}[Q^i,P_i,N]$ describes an ordinary gauge system with a constraint
$P_0= 0$, so that the coordinate $Q^0$ is pure gauge, that is,
$Q^0$ is not associated to a physical degree of freedom. This
coordinate can be defined as an arbitrary function of $\tau$ by
means of a canonical gauge choice. Writing $Q^0$ in terms of $q^i$
and $p_i$ we have a function of the original phase space variables
whose Poisson bracket with $H=P_0$ is positive definite; as $H$
differs from the original Hamiltonian constraint only by a positive
definite function, then we can always define a global phase time as
\begin{equation}
t(q^i,p_i)\equiv Q^0(q^i,p_i)
\end{equation}
because
$
[t(q^i,p_i),H(q^i,p_i)]=[Q^0,P_0]=1,
$
and then
\begin{equation}
[t(q^i,p_i),{\cal H}(q^i,p_i)]>  0.
\end{equation}
The key point that allows to define a global phase time for the
minisuperspace by imposing a canonical gauge condition on the
associated gauge system described by $(Q^i,P_i)$ is that in terms
of these variables we have a natural choice for a function whose
Poisson bracket with the constraint is non vanishing everywhere.

\subsection{Path integral}

The action ${\cal S}[Q^i,P_i,N]$ is stationary when the coordinates
$Q^i$ are fixed at the boundaries. The coordinates and momenta
$(Q^i,P_i)$ describe a gauge system with a linear constraint, so
that this action allows to obtain the amplitude for the transition
$|Q^i_1,\tau_1\rangle \to |Q^i_2,\tau_2\rangle$ by the usual
Faddeev--Popov procedure:
\begin{eqnarray}
\langle Q^i_2,\tau_2|Q^i_1,\tau_1\rangle & = &\nonumber \\
 &  & \hspace{-3.2cm}\int DQ^0 DP_0 DQ^\mu DP_\mu DN
\delta (\chi) |[\chi, P_0]|e^{i{\cal S}[Q^i,P_i,N]}\end{eqnarray}
with
${\cal S}[Q^i,P_i,N]$ the gauge invariant action (\ref{cals}),
and where $\chi=0$ can be any canonical gauge condition. The
Faddeev--Popov determinant $|[\chi, P_0]|$ ensures that the result
does not depend on the gauge choice. If we perform the functional
integration on the lapse $N$ enforcing the paths to lie on the
constraint hypersurface $P_0=0$, we obtain
\begin{eqnarray}
\langle Q^i_2,\tau_2|Q^i_1,\tau_1\rangle   & = &  \int DQ^0  DQ^\mu DP_\mu
\delta (\chi) |[\chi, P_0]|\nonumber\\
& &\hspace{-2.5cm}\times\exp \left(i\int_{\tau_1}^{\tau_2}\left[ P_\mu{dQ^\mu\over d\tau}-
h(Q^\mu,P_\mu,\tau)\right]d\tau  \right),\end{eqnarray}
where $h\equiv\partial f/\partial\tau$ is the true Hamiltonian of the reduced
system.  The path integral gives an amplitude between states characterized by
the variables which,  when fixed at the boundaries, make the action stationary.
As ${\cal S}$ is stationary when the $Q^i$ are fixed, then we  choose the gauge
in the most general form giving $Q^0$ as a function of the other coordinates
$Q^\mu$ and $\tau$; thus a choice of the boundary values of
the physical coordinates and $\tau$ fixes the boundary values of $Q^0$. With a
choice
$
\chi\equiv Q^0-T(Q^\mu,\tau)=0
$
and after trivially integrating on $Q^0$ we finally obtain
\begin{eqnarray}
\langle Q^i_2,\tau_2|Q^i_1,\tau_1\rangle & = & \int  DQ^\mu DP_\mu\nonumber\\
& &\hspace{-2.5cm}
 \mbox{}\times\exp \left( i\int_{\tau_1}^{\tau_2}
\left[ P_\mu{dQ^\mu\over
d\tau}-h(Q^\mu,P_\mu,\tau)\right]d\tau  \right)
    \label{325}
\end{eqnarray}
so that we have $\langle Q^i_2,\tau_2|Q^i_1,\tau_1\rangle  =  \langle
Q^\mu_2,\tau_2|Q^\mu_1,\tau_1\rangle.$ Now, what we are looking for is an amplitude between states
characterized by the original variables of a minisuperspace.
Because the original action $S[q^i,p_i,N]$ is stationary when the
coordinates $q^i$ are fixed at the boundaries, it is usual to look
for a propagator of the form
\begin{equation}
\langle q^i_2|q^i_1\rangle  ,
\end{equation}
so that the states are characterized only by the coordinates. But,
as we have already remarked, in cosmology it is not always possible
to define a time in terms of the $q^i$ only; then the amplitude
$\langle q^i_2|q^i_1\rangle $ could not in general be understood as
the probability that the observables of the system take a certain
values at time $t$ if at a previous time they took other given
values.

If we pretend that
$$\langle Q_2^\mu,\tau_2|Q_1^\mu,\tau_1\rangle= \langle q_2^i|q_1^i\rangle $$ the paths
should be weighted by the action ${\cal S}$ in the same way that
they are weighted by $S$, and the quantum states
$|Q^\mu,\tau\rangle $ should be equivalent to $|q^i\rangle .$ As
the path integral in the variables $(Q^i,P_i)$ is gauge invariant,
this requirement is verified if it is possible to impose a
--globally good-- gauge condition $\tilde \chi=0$ such that
$\tau=\tau(q^i)$ is defined. But this can be fulfilled only in the
case that a global time $t(q^i)$ exists, which is not true in
general. In the most general case a global phase time must
necessarily involve the momenta, and then we cannot fix the gauge
in such a way that $\tau=\tau(q^i)$. Hence, we should admit the
possibility of identifying the quantum states in the original phase
space not by $q^i$ but by a complete set of functions of both the
coordinates and momenta $q^i$ and $p_i$.

This may suggest to give up the idea of obtaining an amplitude for
states characterized by the coordinates. However, while a
deparametrization in terms of the momenta may be completely valid
at the classical level, at the quantum level there is an obstacle
which is peculiar of gravitation \cite{ba93}: There are basically
two representations for quantum operators, the coordinate
representation and the momentum representation, in which the states
are characterized by occupation numbers associated to given values
of the momenta. The last one is appropriate when the theory under
consideration allows for the existence of asymptotically free
states, so that an interpretation in terms of creation and
annihilation operators exists. In quantum cosmology these
asymptotic states do not, in general, exist. The appropriate
representation is then a coordinate one, in which the quantum
states are represented by wave functions in terms of the
coordinates. The usual Dirac--Wheeler--DeWitt quantization with
momentum operators in the coordinate representation follows this
line; but, as we have already noted, this formalism is devoided of
a clear notion of time and evolution, unless a time in terms of
only the canonical coordinates exists.

An intermediate way can then be followed: When the constraint
allows for the existence of an intrinsic time, our
deparametrization and path integral quantization procedure
straightforwardly gives the transition amplitude for states
characterized by the original coordinates; this provides a
quantization with a clear distinction between time and observables.
On the other hand, when only an extrinsic time exists we change
from the original variables $(q^i,p_i)$ to a set $(\tilde
q^i,\tilde p_i)$ defined in such a way that the Hamiltonian
constraint of a given model has a non vanishing potential; then an
intrinsic time exists in terms of the coordinates $\tilde q^i$, and the action
$S[\tilde q^i,\tilde p_i,N]$ is stationary when the $\tilde q^i$
are fixed at the boundaries. Therefore our procedure yields the
transition amplitude for states characterized by the new
coordinates, which is given by
$$\langle \tilde q^i_2|\tilde q^i_1\rangle  =\langle Q_2^\mu,\tau_2|Q_1^\mu,\tau_1\rangle.$$
In both cases we obtain a consistent quantization with a clear
distinction between time and observables. Though this seems to
complicate the interpretation of the resulting propagator, the
original momenta turn to appear only in the time variable, while
the new coordinates corresponding to the physical degrees of
freedom depend on the $q^i$ only (a detailed discussion has been
given in the context of the quantization of the Taub anisotropic
cosmology; see \cite{gisi02,si02b}).

The form of the Hamiltonian $h$ of the reduced system depends on
the choice of the function $f$. We can choose $f$ so that
that the amplitude $\langle Q_2^\mu,\tau_2|Q_1^\mu,\tau_1\rangle $
is equivalent to $\langle \tilde q_2^i|\tilde q_1^i\rangle $. This
requires that the Hamiltonian constraint allows to define a time in
terms of the coordinates $\tilde q^i$ and that the end point terms
vanish on the constraint surface and in the gauge $\tilde \chi=0$
defining $\tau=\tau(\tilde q^i),$ that is,
\begin{equation} \left.\left[ \overline Q^i\overline P_i -W+Q^\mu P_\mu-
f\right]_{\tau_1}^{\tau_2}\right|_{P_0=0,\tilde \chi=0}=0 .
\end{equation}
Because the action ${\cal S}$ is gauge-invariant, this ensures that
with any gauge choice the paths are weighted in the same way by
${\cal S}$ and $S$. This requirement gives a prescription for the
generator $f(\overline Q^\mu,P_\mu,\tau)$ which determines the
reduced Hamiltonian $h=\partial f/\partial\tau$. As $f$ depends
only on observables, $h$ commutes with the complete Hamiltonian
$K=NP_0+h$, so that
$${dh\over d\tau}={\partial^2 f\over \partial \tau^2}.$$
Thus a generator $f$ linear in $\tau$ yields a conserved
Hamiltonian for the reduced system.

The reduced Hamiltonian $h$ could be both positive or
negative-definite. As we shall illustrate with the  second example of the
next section, in  general the sign of $h$ will be in correspondence with the
sign of a non vanishing momentum of the set $\{\tilde p_i\}$ in
terms of which the constraint surface splits into two sheets. The
formalism will therefore include two theories for the physical
degrees of freedom, each one corresponding to each sign of $h$
associated to one of the two sheets of the constraint surface. The
path integral in the reduced space will give two propagators, one
for the evolution of the wave functions of each theory (see
\cite{ba93}, and also \cite{ha86} for an analogous point of view). Note that then, if  our path integral is to be associated to  a canonical quantization,  the splitting of the formulation into two disjoint theories is in correspondence with  two Schr\"{o}dinger equations; so  in general it does not  coincide with the ordinary Wheeler -- DeWitt quantization. However,  the existence of two disjoint theories, one for each sheet of the constraint surface, is a general property resulting from  working with a time $t(\tilde q^i)$, which comes from the fact that we want to identify the path integral in the variables $Q^i$ with a transition amplitude between states given in terms of the coordinates $\tilde q^i$; the nonequivalence between the Schr\"{o}dinger and the Wheeler -- DeWitt quantizations, instead, depends on the model under consideration, and also on the choice of  coordinates. This has  been discussed in detail in  \cite{si03}.

\subsection{Examples}

Consider the Hamiltonian constraint of the most general empty
homogeneous and isotropic cosmological model:
\begin{equation}
{\cal H}=-{\frac{1}{4}}e^{-3\Omega }p _{\Omega }^{2}-ke^{\Omega }+\Lambda
e^{3\Omega }= 0.
\end{equation}
This Hamiltonian corresponds to a universe with arbitrary curvature
$k=-1,0,1 $ and non zero cosmological constant; we shall assume
$\Lambda >0$. If $k=0$ we have the de Sitter universe. The
classical evolution corresponds to an exponential expansion. For
both $k=0$ and $k=-1$ the potential is never zero, and then $p
_{\Omega }$ cannot change its sign. Instead, for the closed model
$p _{\Omega }=0$ is possible.

It is convenient to work with the rescaled Hamiltonian
$H=e^{-\Omega}{\cal H}
:$
\begin{equation}
H=-{\frac{1}{4}}e^{-4\Omega}p_\Omega^2 -k+\Lambda e^{2\Omega}= 0.
\end{equation}
The constraints $H$ and ${\cal H}$ are equivalent because they
differ only in a positive  factor. The $\tau -$independent
Hamilton--Jacobi equation for the Hamiltonian $H$ has the solution
\begin{equation}
W(\Omega ,\overline{P}_{0})=2\eta\int d\Omega e^{2\Omega }%
\sqrt{\Lambda e^{2\Omega }-k-\overline{P}_{0}},
\end{equation}
which is the generating function of the canonical transformation
$(\Omega ,\pi _{\Omega })\to (\overline{Q}^{0},\overline{P}_{0})$
defined by
\begin{equation}
\overline{Q}^{0}=-\eta\Lambda ^{-1}\sqrt{\Lambda e^{2\Omega }-k-\overline{P}_{0}}, \ \ \ \
\ \ \  \overline{P}_{0}=H,
\end{equation}
with $\eta = sgn( p _{\Omega })$. Then we define the function $F=\overline{Q}^{0}P_{0}+f(\tau )$
which generates the second canonical transformation yielding a non
vanishing true Hamiltonian $h=\partial f/\partial \tau $ and
$Q^{0}=\overline{Q}^{0}$, $
\overline{P}_{0}=P_{0}.$

The variables $Q^{0}$ and $P_{0}$ describe the gauge system into
which the model has been turned. The gauge can now be fixed by
means of a $\tau - $dependent canonical condition like $\chi \equiv
Q^{0}-T(\tau )=0$ with $T$ a monotonic function of $\tau $. Then we
can define the time as
\begin{equation}
t=Q^{0}|_{P_{0}=0}=-\eta\Lambda ^{-1}\sqrt{\Lambda
e^{2\Omega }-k},
\end{equation}
or, using the constraint equation,
\begin{equation}
t(\Omega ,p _{\Omega })=-{1\over 2}\Lambda ^{-1}e^{-2\Omega }p _{\Omega },
\label{time}
\end{equation}
which is in agreement with the time obtained by matching the model
with the ideal clock \cite{befe95,desi99a}. An important difference
between the cases $ k=-1$ and $k=1$ arises: for $k=-1$ the
constraint surface splits into two disjoint sheets. In this case
the evolution can be parametrized by a function of the coordinate $
\Omega $ only, the choice given by the sheet on which the system
remains: on the sheet $p _{\Omega }>0$ the time is $t=-\Lambda
^{-1}
\sqrt{\Lambda e^{2\Omega }+1}$, while  on the sheet $p_{\Omega }<0$
we have $t=\Lambda ^{-1}\sqrt{\Lambda e^{2\Omega }+1}$. The
deparametrization of the flat model is completely analogous. For
the closed model, instead, the potential can be zero and the
topology of the constraint surface is no more equivalent to that of
two disjoint planes. Although for $
\Omega =-\ln (\sqrt{\Lambda })$ we have $V(\Omega )=0$ and $p _{\Omega }=0,
$ at this point it is $dp _{\Omega }/d\tau \neq 0$.
Hence in this case  $\Omega $ cannot  parametrize
the evolution, because the system can go from $(\Omega ,p _{\Omega })$ to $
(\Omega ,-p _{\Omega })$; therefore  a  global
phase time must necessarily be defined as a function of both the coordinate and the momentum.

The system has one degree of freedom and one constraint, so that it
is pure gauge. In other words, there is only one physical state:  from a given point in the phase space
any other point on the constraint surface can be reached by means of a finite
gauge transformation. This provides a  proof for the consistency of our
procedure: it should be possible to verify that the transition
probability written in terms of the variables which include a
globally well defined time is equal to unity.

The quantization is straightforward, and the observation above is
reflected in   that we  obtain the propagator \cite{gisi01b}
\begin{equation}
\langle Q_{2}^{0},\tau _{2}|Q_{1}^{0},\tau _{1}\rangle = \exp \left( -
i\int_{\tau _{1}}^{\tau _{2}}{\frac{\partial f}{\partial
\tau }}d\tau \right) ,
\end{equation}
and then the probability for the transition from $Q_{1}^{0}$ at
$\tau _{1}$ to $Q_{2}^{0}$ at $\tau _{2}$ is indeed
\begin{equation}
\left| \langle Q_{2}^{0},\tau _{2}|Q_{1}^{0},\tau _{1}\rangle \right| ^{2}=1.
\end{equation}
When the model is open or flat the coordinates $\Omega $ and
$Q^{0}$ are uniquely related; hence the result simply reflects that
once a gauge is fixed there is only one possible value of the scale
factor $ a\sim e^{\Omega}$ at each $\tau $, and
\be
|\langle\Omega_2|\Omega_1\rangle |^{2}=1.
\ee
But in the case of the closed model, at each $\tau $ there are two
possible values of the coordinate $\Omega $; instead, there is only one possible
value of the momentum $p_{\Omega }$ at each $\tau $. Hence the transition
probability in terms of $ Q^{0}$ does not correspond to the
evolution of the coordinate $\Omega $, but rather of its
derivative, and the amplitude $\langle Q_{2}^{0},\tau
_{2}|Q_{1}^{0},\tau _{1}\rangle $ corresponds to an amplitude
$\langle p
_{\Omega ,2}| p _{\Omega ,1}\rangle $, and we have
\begin{equation}
|\langle p _{\Omega ,2}|p _{\Omega ,1}\rangle |^{2}=1.
\end{equation}
The fact that the resulting amplitude is not equivalent to $\langle
\Omega _{2}|\Omega _{1}\rangle $ is clearly not  a failure
of the quantization procedure, because for this model a characterization of the
states in terms of only the original coordinates is not possible if we
want to retain a formally right notion of time on the whole evolution.

Now let us apply our formulation to a system with true degrees of
freedom; consider a Hamiltonian constraint of the form
\begin{equation}
{\cal  H}=G(\tilde q^2)(\tilde p_1^2-\tilde p_2^2)+V(\tilde q^1 ,\tilde q^2 )
= 0,
\end{equation}
where $G(\tilde q^2 )> 0$. This constraint includes homogeneous and
isotropic models, both relativistic and dilatonic, and also some
anisotropic models, like the Bianchi type I, the Kantowski--Sachs
universe and also the Taub universe (after the appropriate
canonical transformation introduced in \cite{gisi02}). We shall restrict our
analysis to the cases in which the potential $ V(\tilde q^1 ,\tilde
q^2 )$ has a definite sign, so that ${\tilde q^i}$ is a set of
coordinates including a global time; we shall assume $V> 0$. We
shall also suppose that coordinates
\begin{equation}
x=x(\tilde q^1 +\tilde q^2),\ \ \ \ \ \ y=y(\tilde q^1 -\tilde
q^2)\end{equation}
can be introduced  so that $4(\partial x/\partial\tilde q^1)(\partial y/
\partial\tilde q^1)=V/ G$; then we can write the constraint in the (scaled)
equivalent form
 \begin{equation}
H=p_x p_y+ 1= 0.\label{424}
\end{equation}
The solution of the corresponding Hamilton--Jacobi equation can be
chosen so that the canonical variables of the associated gauge
system are given by
\begin{eqnarray}
Q^0 & = & {y\over P},\nonumber\\
 Q & = & x+{1\over P^2}\left(y(1- P_0)-\eta T(\tau)\right),\nonumber\\
P_0 & = & p_x p_y+1,\nonumber\\
 P & = &p_x.\label{426}
\end{eqnarray}
Thus a canonical gauge condition $\chi\equiv Q^0-T(\tau)=0$ is
associated with the extrinsic time
$
t= y/p_x.$ We can also define an intrinsic time, which is
related with the obtention of a transition amplitude between states
characterized by the coordinates. The end point terms associated to
the canonical transformation $(x,y,p_x,p_y)\to (Q^i,P_i)$ are of
the form
\be B(\tau)  =  2Q^0-Q^0P_0-2\eta{T(\tau)\over P}.
\ee
On the constraint surface $P_0=0$
these terms clearly vanish in  gauge
$
\chi\equiv \eta Q^0P-T(\tau)=0
$
which is in correspondence with the intrinsic time and true
Hamiltonian(s)
\be
t(\tilde q^1,\tilde q^2)=\eta\, y(\tilde q^1-\tilde q^2),\ \ \ \ \
\  h(Q,P,\tau)={\eta\over P}{dT\over d\tau},\ee
with $\eta = sgn
(p_x)=sgn (\tilde p_1+\tilde p_2)= sgn (\tilde p_2)$, because $V>0$
ensures that $|\tilde p_2|>|\tilde p_1|$. The propagator for the
transition $|\tilde q^1_1,\tilde q^2_1\rangle \to |\tilde
q^1_2,\tilde q^2_2\rangle $ is given by
\begin{eqnarray}
\langle\tilde q^1_2,\tilde q^2_2\vert\tilde q^1_1,\tilde q^2_1 \rangle & = &\nonumber\\
& & \hspace{-2cm}\int
DQDP\exp\left[i\int_{T_1} ^{T_2}
\left(PdQ-{\eta\over P}dT\right) \right],\label{437}
\end{eqnarray}
where the end points are given by
$T_1=\pm y(\tilde q^1_1-\tilde q^2_1)$ and $T_2=\pm y(\tilde q^1_2-\tilde
q^2_2).$ Note that with the gauge choice defining an intrinsic time, the
observable $Q$ reduces to a function of only the original coordinates:
$$Q|_{\chi=0}=x(\tilde q^1+\tilde q^2).$$
Hence the paths go from $Q_1=x(\tilde q^1_1+\tilde q^2_1)$ to
$Q_2=x(\tilde q^1_2+\tilde q^2_2)$. The propagator in the reduced
space is therefore that of a system with a true degree of freedom
given by the coordinate $Q$. Also, because $\tilde p_2$ does not
vanish on the constraint surface, the coordinate $\tilde q^2$ is
itself a global time, namely $t^*$; hence, though $\tilde q^2$ is
not the time parameter in the path integral, the transition
amplitude could be written as $\langle x_2,t^*_2\vert x_1, t^*_1
\rangle $. Observe that by considering both possible signs of the
reduced Hamiltonian, this path integral gives the transition
amplitude for both theories corresponding to both sheets of the
constraint surface identified by the sign of the momentum
$\tilde p_2.$

\section{A closed universe as a system without asymptotic states
and the problem of time}

Several years ago a new approach to constructing quantum
geometrodynamics was proposed by Savchenko, Shestakova and
Vereshkov \cite{ssv99, ssv00, shest99, ssv01a, ssv01b}. The central
part in this approach is given to a Schr\"odinger equation for a
wave function of the Universe which contains time as an external
parameter like in ordinary quantum mechanics. However, the
appearance of time in the Schr\"odinger equation is a consequence
of breaking down gauge invariance of the theory. The proposed
formulation is radically distinguished from the generally accepted
Wheeler--DeWitt quantum geometrodynamics, so one needs to have a
strong grounds for justifying this formulation.

\subsection{Asymptotic states and gauge invariance}

A key point of the authors' argumentation is the analysis of the
role of asymptotic states in quantum gravity \cite{ssv01a}. It is
emphasized that any gauge-invariant quantum field theory is
essentially based on the assumption about asymptotic states.
Indeed, in the case of canonical quantization, in order to separate
true physical degrees of freedom from ``nonphysical'' ones we
need to resolve gravitational constraints. It can be done in the
limits of perturbation theory in asymptotically flat spaces or in
some special cases. But in a general situation, if the Universe has
some nontrivial topology and does not possess asymptotic states,
this procedure meets insurmountable mathematical difficulties.

In the path integral approach, which was accepted by the authors
as most adequate, asymptotic boundary conditions ensure the
BRST-invariance of a path integral and play the role of selection
rules; as a consequence, the path integral does not depend on a
gauge-fixing function (see \cite{hen85}). Since a closed universe
is a system without asymptotic states, it is not correct in this case
to impose asymptotic boundary conditions in a path integral, so that
the set of all possible transition amplitudes determined through the
path integral inevitably involves gauge-noninvariant ones.

If the path integral is considered without asymptotic boundary
conditions it should be skeletonized on a {\it full} set of
gauge-noninvariant equations obtained by varying an appropriate
effective action including ghost and gauge-fixing terms.
Further, there are two nonequivalent ways to proceed: to make
use of the Batalin--Vilkovisky (Lagrangian) \cite{bv81}
or the Batalin--Fradkin--Vilkovisky (Hamiltonian)
\cite{fv75, bv77, ff78} effective action. There exists
the difference in the structure of ghost sectors, which, in turn,
results from the fact that the gauge group of gravity does not
coincide with the group of canonical transformations generated by
gravitational constraints. Two formulations based on Lagrangian and
Hamiltonian effective actions could be done equivalent in the
gauge-invariant sector, the latter being singled out by means of
asymptotic boundary conditions. Again, here we can see a crucial
role of the assumption about asymptotic states in ensuring gauge
invariance. In the situation without asymptotic states one has to
make a choice between these two effective actions; the authors
give preference to the Lagrangian formalism since it maintains the
original group of gauge transformations. Moreover, one cannot
ensure the BRST-invariance of the action without imposing
asymptotic boundary condition, and the BFV scheme is broken then.

This approach leads to the extended set of Einstein equations in
which the constraints are broken already at the classical level.
Eventually, this causes the dynamical Schr\"odinger equation and
the appearance of time. A similar modification of the Hamiltonian
constraint and the related time-dependent Schr\"odinger equation
was discussed early by Weinberg \cite{we89} and Unruh
\cite{unruh89a}. The modification aimed to solve the cosmological
constant problem, and the cosmological constant appeared to be a
Hamiltonian eigenvalue. It resulted from an additional condition on
a metric tensor which did not fix a gauge. It is clear then, that
the modification suggested by Weinberg and Unruh did not touch
other gravitational constraints and equations of motion and was
considered as a remedy for a particular (though very important)
problem.

Another argument in favor of the gauge-noninvariant approach is
the parametrization noninvariance of the Wheeler--DeWitt
equation \cite{hp86, hal88}. The authors consider a unified
interpretation of the choice of gauge variables (parametrization)
and the choice of gauge conditions; the latter ones together
determine equations for the metric components $g_{0\mu}$, fixing a
reference frame, as it is illustrated by the scheme \cite{shest00}:

\begin{center}
\begin{tabular}{ccc}
Parametrization & & \\
$g_{0\mu}=v_{\mu}\left(\mu_{\nu},\gamma_{ij}\right)$ & & \\
+ &  $\Rightarrow$ & Equations for $g_{0\mu}$\\
Gauge conditions & &
  $g_{0\mu}=v_{\mu}\left(f_{\nu}\left(\gamma_{ij}\right),\gamma_{ij}\right)$\\
$\mu_{\nu}=f_{\nu}\left(\gamma_{ij}\right)$ & &
\end{tabular}
\end{center}

\noindent
Here $\mu_{\nu}$ are new gauge variables, in particular, the lapse and
shift functions, $N$ and $N_i$, $\gamma_{ij}$ is 3-metric.
Thus even if one considers $\mu_{\nu}$ as independent of $\gamma_{ij}$,
different parametrizations will correspond to different reference frames.
This leads to the conclusion that  a transition to another gauge
variable is formally equivalent to imposing a new gauge condition, and vice
versa, and the parametrization noninvariance of the Wheeler--DeWitt
equation is ill-hidden gauge noninvariance.

\subsection{Hamiltonian dynamics in extended phase space}

After these preliminary notes let us go into mathematical details.
The authors consider a simple minisuperspace model with the gauged action
\begin{eqnarray}
\label{action}
S & = & \!\int\!dt\,\biggl\{
  \displaystyle\frac12 v(\mu, Q^a)\gamma_{ab}\dot{Q}^a\dot{Q}^b
  -\frac1{v(\mu, Q^a)}U(Q^a)\nonumber\\
  & & \mbox{}+\pi\left(\dot\mu-f_{,a}\dot{Q}^a\right)
  -i w(\mu, Q^a)\dot{\bar\theta}\dot\theta\biggr\}.
\end{eqnarray}
Here $Q^a$ stands for physical variables such as a scale factor or
gravitational-wave degrees of freedom and material fields, and
an arbitrary parametrization of a gauge variable $\mu$ determined by
the function $v(\mu, Q^a)$ is accepted. For example, in the case
of isotropic universe or the Bianchi IX model $\mu$ is bound to the
scale factor $r$ and the lapse function $N$ by the relation
\begin{equation}
\label{paramet}
\displaystyle\frac{r^3}{N}=v(\mu, Q^a).
\end{equation}
The special class of gauges not depending on time is used
\begin{equation}
\label{frame_A}
\mu=f(Q^a)+k;\quad
k={\rm const}.
\end{equation}
It is convenient to present the gauge in a differential form,
\begin{equation}
\label{diff_form}
\dot{\mu}=f_{,a}\dot{Q}^a,\quad
f_{,a}\stackrel{def}{=}\frac{\partial f}{\partial Q^a}.
\end{equation}
$\theta,\,\bar\theta$ are
the Faddeev--Popov ghosts after replacement
$\bar\theta\to -i\bar\theta$. Further,
\begin{equation}
\label{w_def}
w(\mu, Q^a)=\frac{v(\mu, Q^a)}{v_{,\mu}};\quad
v_{,\mu}\stackrel{def}{=}\frac{\partial v}{\partial\mu}.
\end{equation}
Varying the effective action (\ref{action}) with respect to
$Q^a$,$\mu$,$\pi$ and $\theta$,$\bar\theta$ one gets,
correspondingly, motion equations for physical variables, the
constraint, the gauge condition and equations for ghosts.
The extended set of Lagrangian equations is complete in
the sense that it enables one to formulate the Cauchy problem. The
explicit substitution of trivial solutions for ghosts and the
Lagrangian multiplier $\pi$ to this set of equations turns one
back to the gauge-invariant classical Einstein equations.

The path integral approach does not require the construction of a
Hamiltonian formulation before deriving the Schr\"odinger
equation, but it implies that the Hamiltonian formulation can be
constructed. Indeed, in the class of gauges (\ref{diff_form}) the
Hamiltonian can be obtained in a usual way, according to the rule
$H=p\dot q-L$, where $(p,q)$ are the canonical pairs of extended
phase space (EPS), by introducing momenta conjugate to all degrees
of freedom including the gauge ones,
\begin{eqnarray}
\label{Hamilt}
H & = & P_a\dot Q^a+\pi\dot\mu+\bar\rho\dot\theta+\dot{\bar\theta}\rho-L
 \nonumber\\
 & = & \frac12G^{\alpha\beta}P_{\alpha}P_{\beta} +\frac1{v(\mu,Q^a)}U(Q^a)
 -\frac i{w(\mu,Q^a)}\bar\rho\rho,\nonumber\\
 &  &
\end{eqnarray}
where $\alpha=(0,a),\;Q^0=\mu$,
\begin{equation}
\label{Galphabeta}
G^{\alpha\beta}=\frac1{v(\mu,Q^a)}\left(
\begin{array}{cc}
f_{,a}f^{,a}&f^{,a}\\
f^{,a}&\gamma^{ab}
\end{array}
\right).
\end{equation}
The system of Hamiltonian equations in EPS
\begin{equation}
\label{Xequat}
\dot X=\{H,X\},\quad
X=(P_a,\;Q^a,\;\pi,\;\mu,\;\bar\rho,\;\theta,\;\rho,\;\bar\theta),
\end{equation}
is completely equivalent to the extended set of Lagrangian
equations, the constraint and the gauge condition acquiring the
status of Hamiltonian equations. The idea of extended phase space
is exploited in the sense that gauge and ghost degrees of freedom
are treated on an equal basis with other variables. This gave rise
to the name ``quantum geometrodynamics in extended phase space''
accepted by the authors.

Obviously, the Hamiltonian dynamics is constructed here in a
different way than in the BFV approach where constraints are
maintained and play a central role. Here the Hamiltonian constraint
is modified and looks like follows:
\begin{eqnarray}
\label{pi-Ham}
\dot\pi & = & -\frac i{w^2(\mu,Q^a)}w_{,\mu}\bar\rho\rho
  +\frac{1}{v^2(\mu,Q^a)}v_{,\mu}\times\nonumber\\
& & \hspace{-1cm}\mbox{}\times\left[\frac12\left(P_aP^a
   +2\pi f_{,a}P^a+\pi^2f_{,a}f^{,a}\right)
+U(Q^a)\right].
\end{eqnarray}
The gauge-dependent terms can be eliminated making use of trivial
solutions for $\pi$ and ghosts. It is generally accepted that
these trivial solutions can be singled out by asymptotic boundary
conditions, if one ignores the problem of Gribov's copies.
Further, the restriction on the class of admissible
parametrization, $v(\mu,Q)=\frac{u(Q)}{\mu}$, together with the
trivial solutions for $\pi$ and ghosts reduces the constraint to
the form
\begin{equation}
\label{Dir-constr}
{\cal T}=\frac1{2u(Q^a)}P_aP^a+\frac1{u(Q^a)}U(Q^a
)=0.
\end{equation}
The restriction on the class of parametrization is necessary for
the physical Hamiltonian be proportional to the constraint,
$H_0=\mu{\cal T}$. We come to the conclusion that the primary and
secondary constraints $\pi=0,\quad {\cal T}=0$ correspond to a
particular situation when it is possible to single out the trivial
solutions for $\pi$ and ghosts by asymptotic boundary conditions.
So, in this sense, the Dirac quantization is applicable to
systems with asymptotic states only. The same is true for the BFV
approach which inherits most of features of the Dirac
quantization.

\subsection{Quantum geometrodynamics in extended phase space}

The constraint (\ref{pi-Ham}) can be presented in the form $H=E$,
where $E$ is a conserved quantity (a new integral of motion).
Correspondingly, in a quantum theory the relation $H=E$ should be
replaced by a stationary Schr\"odinger equation,
$H|\Psi\rangle=E|\Psi\rangle$, the Hamiltonian spectrum in EPS being
not limited by the unique zero eigenvalue.

So, there is no reason to require that a wave function of a closed
universe should satisfy the Wheeler--DeWitt equation. Independently
of our notion of gauge invariance or noninvariance of the theory,
the wave function should obey some Schr\"odinger equation.
The Schr\"odinger equation is derived from the path integral with
the effective action (\ref{action}) by a standard method originated
by Feynman \cite{fey48, cheng}. For the present model it reads
\begin{equation}
\label{SE1}
i\,\frac{\partial\Psi(\mu,Q^a,\theta,\bar\theta;\,t)}{\partial t}
 =H\Psi(\mu,\,Q^a,\,\theta,\,\bar\theta;\,t),
\end{equation}
where
\begin{eqnarray}
\label{H}
H& = &-\frac i w\frac{\partial}{\partial\theta}
   \frac{\partial}{\partial\bar\theta}
  -\frac1{2M}\frac{\partial}{\partial Q^{\alpha}}MG^{\alpha\beta}
   \frac{\partial}{\partial Q^{\beta}}\nonumber\\
& &\mbox{}
  +\frac1v(U-V);
\end{eqnarray}
the operator $H$ corresponds to the Hamiltonian in EPS
(\ref{Hamilt}). $M$ is the measure in the path integral,
\begin{equation}
\label{M}
M(\mu, Q^a)=v^{\frac K2}(\mu, Q^a)w^{-1}(\mu, Q^a);
\end{equation}
$K$ is a number of physical degrees of freedom; the wave function is defined
on extended configurational space with the coordinates
$\mu,\,Q^a,\,\theta,\,\bar\theta$.
$V$ is a quantum correction to the potential $U$, that depends on the chosen
parametrization (\ref{paramet}) and gauge (\ref{frame_A}):
\begin{eqnarray}
V&=&\frac5{12w^2}\left(w^2_{,\mu}f_{,a}f^{,a}+2w_{,\mu}f_{,a}w^{,a}
    +w_{,a}w^{,a}\right)+\nonumber\\
& &\hspace{-1.3cm}\mbox{}+\frac1{3w}\left(w_{,\mu,\mu}f_{,a}f^{,a}+2w_{,\mu,a}f^{,a}
    +w_{,\mu}f_{,a}^{,a}+w_{,a}^{,a}\right)+\nonumber\\
& &\hspace{-1.3cm}\mbox{}+\frac{K-2}{6vw}\left(v_{,\mu}w_{,\mu}f_{,a}f^{,a}+v_{,\mu}f_{,a}w^{,a}+\right.
   \nonumber\\
& &\hspace{-1.3cm}\mbox{} +w_{,\mu}\left.f_{,a}v^{,a}+v_{,a}w^{,a}\right)-\nonumber\\
& &\hspace{-1.3cm}\mbox{}-\frac{K^2-7K+6}{24v^2}\left(v^2_{,\mu}f_{,a}f^{,a}+2v_{,\mu}f_{,a}v^{,a}
    +v_{,a}v^{,a}\right)+\nonumber\\
\label{V}
& &\hspace{-1.3cm}\mbox{}+\frac{1-K}{6v}\left(v_{,\mu,\mu}f_{,a}f^{,a}+2v_{,\mu,a}f^{,a}
    +v_{,\mu}f_{,a}^{,a}+v_{,a}^{,a}\right).
\end{eqnarray}

Let us emphasize that the Schr\"odinger equation (\ref{SE1}) -- (\ref{V})
is {\it a direct mathematical consequence} of a path integral with
the effective action (\ref{action}) without asymptotic boundary
conditions. Once we reject imposing asymptotic boundary conditions,
we {\it are doomed} to come to a gauge-dependent description of the Universe.

The general solution to the Schr\"odinger equation has the following structure:
\begin{eqnarray}
\label{GS-A}
\Psi(\mu,\,Q^a,\,\theta,\,\bar\theta;\,t)
 & = & \nonumber\\
& & \hspace{-2.8cm}\int\Psi_k(Q^a,\,t)\,\delta(\mu-f(Q^a)-k)\,(\bar\theta+i\theta)\,dk.
\end{eqnarray}

As one can see, the general solution is a superposition of
eigenstates of a gauge operator,
\begin{eqnarray}
\label{k-vector}
\{\mu-f(Q^a)\}|k\rangle = k\,|k\rangle;& &\nonumber\\
|k\rangle = \delta\left(\mu-f(Q^a)-k\right).& &
\end{eqnarray}
It can be interpreted in the spirit of Everett's ``relative state''
formulation. In fact, each element of the superposition (\ref{GS-A})
describes a state in which the only gauge degree of freedom $\mu$ is
definite, so that time scale is determined by processes
in the physical subsystem through functions
$v(\mu,\,Q^a),\,f(Q^a)$ (see (\ref{paramet}), (\ref{frame_A})),
while $k$ being determined by initial clock setting.
The function $\Psi_k(Q^a,\,t)$ describes a state of the physical
subsystem for a reference frame fixed by the condition
(\ref{frame_A}). It is a solution to the equation
\begin{equation}
\label{phys.SE}
i\,\frac{\partial\Psi_k(Q^a;\,t)}{\partial t}
 =H_{(phys)}[f]\Psi_k(Q^a;\,t),
\end{equation}
\begin{eqnarray}
\label{phys.H-A}
H_{(phys)}[f] & = & \nonumber\\
& & \hspace{-2.3cm}\left.\left[-\frac1{2M}\frac{\partial}{\partial Q^a}
  \frac1v M\gamma^{ab}\frac{\partial}{\partial Q^b}
 +\frac1v (U-V)\right]\right|_{\mu=f(Q^a)+k}.\nonumber\\
& &
\end{eqnarray}

One can seek the solution to Eq.(\ref{phys.SE}) in the form
of superposition of stationary state eigenfunctions:
\begin{eqnarray}
\label{stat.states}
\Psi_k(Q^a,\,t)= \sum_n c_{kn}\psi_n(Q^a)\exp(-iE_n t);& &\nonumber\\
H_{(phys)}[f]\psi_n(Q^a) = E_n\psi_n(Q^a).& &
\end{eqnarray}

The eigenvalues $E_n$ should not be associated with energy of any
material field. It results from fixing a gauge condition and
characterizes a subsystem which corresponds to observation means
--a reference frame (see \cite{ssv01a, ssv01b} for details).

Having constructed a general solution to the Schr\"od\-inger
equation one can pose the question if a physical part of the wave
function  could obey the Wheeler--DeWitt equation under some
additional conditions. A natural additional condition in EPS is
the requirement of BRST invariance of the wave function. Indeed,
in the BFV approach the requirement of the BRST invariance leads
immediately to the Wheeler--DeWitt equation. The BRST charge has
an especially simple form for the present model,
\begin{equation}
\label{Omega_BFV}
\Omega_{BFV}=\eta^{\alpha}{\cal G}_{\alpha}={\cal
T}\theta-i\pi\rho,
\end{equation}
where ${\cal G}_{\alpha}=(\pi,\;{\cal T})$ is the full set of
constraints, and due to arbitrariness of BFV ghosts
$\{\eta^{\alpha}\}$ one gets the Wheeler -- DeWitt equation
${\cal T}\,|\Psi\rangle=0$ from the requirement
$\Omega_{BFV}\,|\Psi\rangle=0$.

It is not the case in the approach considered above. We should
remind that the original group of transformations was the group of
gauge transformations in the Lagrangian formalism. It is the reason
why transformations generated by (\ref{Omega_BFV}) do not coincide
with those under which the action (\ref{action}) is invariant. The
BRST charge constructed accordingly the BFV prescription turns out to
be irrelevant in this consideration. Instead there exists another
quantity that plays the role of the BRST generator,
\begin{equation}
\label{BRSTgen}
\Omega
 =w(Q^a,\mu)\;\pi\dot\theta-H\theta
 =-\;i\;\pi\rho-H\theta.
\end{equation}
It is easy to check that (\ref{BRSTgen}) generates transformations
in EPS which are identical to the BRST transformations in the Lagrangian
formalism. Nevertheless, it cannot be presented as a combination
of constraints with infinitesimal parameters replaced by ghosts
and cannot help us to obtain the Wheeler--DeWitt equation
\cite{ssv01b}.

On the other hand, as the authors show, the fact that the wave
function obeys the Wheeler--DeWitt equation does not mean that
this wave function describes the Universe in a gauge invariant
way, i. e. independently of a reference frame. If one puts
$\mu=k,\;E=0$ and restricts the class of parametrizations as
was done above (see (\ref{Dir-constr}) and the text before)
the equation for the physical part of the wave function
$H_{(phys)}\Psi_k(Q^a)=E\Psi_k(Q^a)$
is reduced to the Wheeler--DeWitt equation
with its parametrization noninvariance and without any visible
vestige of a gauge. By construction, however, a solution to this
equation corresponds to a particular choice of a gauge condition
and a particular line in the Hamiltonian spectrum. It is enough
then to fix parametrization to complete the choice of a reference
frame. It confirms the conclusion about ill-hidden
gauge-noninvariance of the Wheeler--DeWitt equation which
has been done in the beginning of this section.

All the above demonstrates that this attempt to derive a gauge
invariant quantum theory from a more general gauge noninvariant
one arises many questions. For a system with asymptotic
states we have the BFV approach where we consider constraints yet
at the classical level before quantization. But even in this case
making use of asymptotic boundary conditions to exclude
gauge-noninvariant terms is an idealization in the sense that we
neglect the problem of Gribov's copies.

\subsection{Topology of the Universe and the irreversibility of time}

In conclusion we shall touch on one of consequences of the
presented approach -- {\it the irreversibility} of a transition to
another reference system in the framework of gauge-noninvariant
description \cite{shest02}. Since the reference frame was declared
to be a constituent of an integrated system as well as a physical
Universe and plays a role of a measuring device, any change in
respect of the reference frame would cause changes in an observed
physical picture. Indeed, let us consider a small variation of the
gauge-fixing function $f(Q^a)$, so that the reference frame will be
fixed by the condition
\begin{equation}
\label{frame_B}
\mu=f(Q^a)+\delta f(Q^a)+k.
\end{equation}
Then, in a new basis corresponding to this reference frame the
wave function will take the form
\begin{eqnarray}
\label{GS-B}
\Psi(\mu,\,Q^a,\,\theta,\,\bar\theta;\,t)
& = & \nonumber\\
& & \hspace{-3.5cm}\int\tilde\Psi_k(Q^a,\,t)\,\delta(\mu-f(Q^a)-\delta f(Q^a)-k)\,
  (\bar\theta+i\theta)\,dk. \nonumber\\
& &
\end{eqnarray}
Here the function $\tilde\Psi_k(Q^a,\,t)$ satisfies Eq.(\ref{phys.SE})
with a Hamiltonian
\begin{eqnarray}
H_{(phys)}[f+\delta f] & = &
\left[-\frac1{2M}\frac{\partial}{\partial Q^a}
  \left(\frac1v M\gamma^{ab}\frac{\partial}{\partial Q^b}\right)\right.
 +\nonumber\\
& &\hspace{-1.5cm}\left.\left. \mbox{}+\frac1v (U-V)\right]\right|_{\mu=f(Q^a)+\delta f(Q^a)+k}.
\label{phys.H-B}\end{eqnarray}
It is obvious that the equation for the physical part of the wave
function with the Hamiltonian (\ref{phys.H-B}) cannot be reduced in
general to the equation with the Hamiltonian (\ref{phys.H-A}). The
measure in the subspace of physical degrees of freedom also depends
on a chosen gauge condition as it follows from the normalization
equation
\begin{eqnarray*}
\int\Psi^*_{k'}(Q^a,\,t)\,\Psi_k(Q^a,\,t)\, \delta(\mu-f(Q^a)-k')& &\nonumber\\
& &\hspace{-6.8cm}\mbox{}\times\delta(\mu-f(Q^a)-k)\,
 dk'\,dk\,M(\mu,\,Q^a)\,d\mu\,\prod_adQ^a=\end{eqnarray*}
\vskip-.8cm
\begin{eqnarray}\label{Psi_norm}
\int\Psi^*_k(Q^a,\,t)\,\Psi_k(Q^a,\,t)\,
 M(f(Q^a)+k,\,Q^a)& &\nonumber\\
& &\hspace{-2.7cm}\mbox{}\times\prod_adQ^a\,dk=1.\end{eqnarray}
Due to smallness of $\delta f(Q^a)$ one can write
\begin{equation}
\label{phys.H-B1}
H_{(phys)}[f+\delta f]=H_{(phys)}[f]+W[\delta f]+V_1[\delta f]
\end{equation}
For our minisuperspace model the operator $W[\delta f]$ reads
\begin{eqnarray}
W[\delta f]&=&\left[\frac1{2M^2}\frac{\partial M}{\partial\mu}\delta f
 \frac{\partial}{\partial Q^a}\left(\frac1v M\gamma^{ab}
 \frac{\partial}{\partial Q^b}\right)\right.-\nonumber\\
\label{W}
& &\mbox{}-\frac1{2M}\frac{\partial}{\partial Q^a}
 \left(\left(\frac1v\frac{\partial M}{\partial\mu}
  -\frac M{v^2}\frac{\partial v}{\partial\mu}\right)\right.\nonumber\\
& &\hspace{-.1cm}\left.\left.\left.\mbox{}\times \delta f\gamma^{ab}\frac{\partial}{\partial Q^b}\right)\right]
 \right|_{\mu=f(Q^a)+k},
\end{eqnarray}
and $V_1[\delta f]$ is the change of quantum potential $V$ (\ref{V})
in first order of $\delta f$.

One can inquire how the probabilities of stationary states (\ref{stat.states})
change under the perturbation $W[\delta f]+V_1[\delta f]$,
which is due to a small variation of the gauged-fixing function $f(Q^a)$.
The Hamiltonian (\ref{phys.H-B}) is Hermitian by construction in
a space with the measure $M(f(Q^a)+\delta f(Q^a)+k,\,Q^a)$,
however it is not Hermitian in a space with the measure
$M(f(Q^a)+k,\,Q^a)$ in which the functions (\ref{stat.states}) are
normalized. In this space the operator (\ref{W}) will
have, in general, anti-Hermitian part. So any transition to
another reference frame must be {\it irreversible}.

This is true for a transition to another reference frame in the
same spacetime region, and this is also true if spacetime consists
of several regions, different reference frames being introduced in
these regions. A nontrivial topology of the Universe may be a reason
why one has to introduce various reference frames in different
spacetime regions. In particular, we can consider mutually
intersected spacetime regions ordered in time. Every time when we
move from one region to another, the physical part of the wave
function would be undergone a non-unitary transformation followed
by changing the measure in the subspace of physical degrees of
freedom that may lead to irreversible consequences in the physical
picture of the Universe. If so, taking into account interaction
with the reference frame --the measuring instrument representing
the observer in quantum theory of gravity-- not only
enables one to introduce time into quantum geometrodynamics, but
also may attach an irreversible character to cosmological evolution.

\section{Discussion}

Among many attempts to give a solution to the problem of time we
have paid a considerable attention to two approaches, which are, as
a matter of fact, very different. The first one, by Simeone and
collaborators, considered in Section 2, is a development of the
unitary approach to quantum gravity inspired by earlier works of
Barvinsky and H\'aj\'{\i}cek. The key point here is a reduction of
the gravitational action to that of an ordinary gauge system. Since
the Hamiltonian constraint is quadratic in the momenta, we come, in
general, to two formulations of the theory which correspond to two
disjoint sheets of the constraint surface given by the two signs of
the momentum conjugated to a time variable. The proposed procedure
enables one to formulate the theory in terms of true degrees of
freedom and then return to a transition amplitude between states
characterized by original variables of phase space, so that the
whole scheme is gauge-invariant. This approach demonstrates how time
can be introduced into the theory {\it without breaking down its
gauge invariance}. Let us emphasize that the requirement of gauge
invariance is conventionally thought to be one of basic
requirements for a physical theory.

In this sense the second approach, presented in Section 3, is very
radical. According to the analysis by Isham \cite{isham92},
approaches to the problem of time can be subdivided into three main
categories: those in which time is identified before quantizing,
approaches in which time is identified after quantizing and
approaches in which time plays no fundamental role at all. The
proposal by Savchenko, Shestakova and Vereshkov does not belong to
any of these categories. In their scheme time naturally appears
while quantizing a gravitational system, namely, while driving a
Schr\"odinger equation from the path integral. In this
consideration time has a status of an external parameter as in
ordinary quantum mechanics. The price for it is a refusal from
gauge invariance of the theory.

We would note that the fact that the Universe does not possess
asymptotic states has not been analysed early from the viewpoint
of its connection with gauge invariance. Traditionally the
Universe was quantized as any gauge system with which we deal in
laboratory physics. On the other hand, in modern field theory one
can find indications that the role of gauge degrees of freedom may
not be just auxiliary. It is enough to mention the Aharanov -- Bohm
effect and instanton solutions. All of them originate from
nontrivial topological structure of spacetime. A future
development will give an objective appraisal to the proposed
approaches to the problem of time which remains to be a
fundamental problem of constructing quantum gravity.

\end{document}